# Sirius and the project of the megalithic enclosures at Gobekli Tepe

Giulio Magli
Faculty of Civil Architecture - Politecnico di Milano
Piazza Leonardo da Vinci 32, 20133 Milan, Italy
Giulio.Magli@polimi.it

The megalithic enclosures of Gobekli Tepe (Urfa, Turkey) are the most ancient stone-built sacred structures known so far, dating back to the $10^{th}$ millennium BC. The possible presence of astronomical targets for these structures is analysed, and it turns out that they may have been oriented – or even originally constructed - to "celebrate" and successively follow the appearance of a "new", extremely brilliant star in the southern skies: Sirius.

## 1. Introduction

Gobekli Tepe, a hill in the province of Urfa in south-eastern Turkey, is the first "temple" about which we are aware of (Schmidt 1998,2001,2006,2010). It is composed by a series of circular enclosures (of which only a few are excavated yet) whose project is based on a megalithic "unit": a huge T-shaped pillar, usually finely engraved. Two such pillars stand in the centre, parallel to each-other, while a series of other pillars is placed around the contour of the circle. Most carvings represent dangerous or anyhow wild animals: felines, foxes, boars, vultures, spiders, snakes, and scorpions. The site is dated to the so called pre-pottery Neolithic (PPN), and was in use between the 10th and the 9th millennium BC; successively, it was left abandoned and intentionally obliterated. Gobekli Tepe is not unique as other Neolithic sites with T-pillars have been documented (see e.g. Celik 2001).

The antiquity of this sacred place is so astonishing that it is extremely difficult to put forward hypotheses on the religion and on the cults which were practised there. In recent years, however, Archaeoastronomy – used with due caution - has proved to be a quite powerful tool in gaining better understanding of the symbolic world of many ancient cultures (see e.g. Ruggles 2005, Magli 2009). In this paper, a preliminary analysis of possible astronomical references at Gobekli Tepe is therefore attempted.

## 2. Discussion

Astronomy is a familiar presence in megalithic sites (although its role has been sometimes stretched too far in the past). Interestingly, and although it may seem strange, there is no doubt that the ancient places which bear the most striking similarities with Gobekli Tepe are the astronomically oriented sanctuaries of Menorca, built some 8000 years *later*. These are oval-shaped enclosures centred on a huge, T-shaped object (commonly referred to as a Taula) composed by a huge pillar and a transverse capital. Such sanctuaries were very probably oriented to the brilliant stars of the southern sky, those of the Crux-Centaurus group, which were slowly disappearing from the Mediterranean sky due to precession (Hoskin 2001). Is it possible that a stellar reference was present at Gobekli as well?

As far as the present author is aware, there is no scholarly work published on the archaeoastronomy of the GT enclosures, while a rectangular building there has been identified as being orientated to the cardinal points (Belmonte and García 2010). On the enclosures there are, however, two non-scholarly publication which are worth considering (Schoch 2012, Collins 2013), In the first, a possible role of the rising of the stars of Orion's belt to the south-east is proposed, while in the second the opposite orientation, identified as targeting the setting of Deneb and of the Cygnus constellation is put forward. Bot these analyses are, in my view, not convincing. Orion indeed would lead to too a high dating for the structures. Regarding a northern orientation, it is in a sense unnatural, as the enclosures are rather open to the south-east; all orientations of "shrines" we are aware of are *from* inside to outside (a few examples: the temples of Malta, the Taulas of Menorca, the Greek temples, the Christian Churches…). Further, there actually exists another possible stellar target which seems to have been overlooked so far. In fact, simulating the sky in the $10^{th}$ millennium BC, it is possible to see that a quite spectacular phenomenon occurred at Gobekli Tepe in that period: the "birth" of a "new" star, and certainly not of an ordinary one, as it is the brightest star and the $4^{th}$ most brilliant object of the sky: Sirius. Indeed precession, at the latitude of Gobekli Tepe, brought Sirius under the horizon in the years around 15000 BC. After reaching the minimum, Sirius started to come closer to the horizon and it became visible again, very low and close to due south, towards 9300 BC.

To check if the enclosures might have been aligned with Sirius, I will consider here the 3 adjacent structures labelled D,C, and B, which are virtually intact and also extremely similar in conception. I stress that the analysis presented below is based on existing maps and satellite images. It must, therefore, be considered as preliminary; a complete theodolite survey of the site and of the horizon is certainly needed to draw more reliable conclusions. Having said that, the extrapolated mean azimuths of the structures (taken as the mid-lines between the two central monoliths) are estimated as follows:

Structure D 172°
Structure C 165°
Structure B 159°

As Sirius is a negative magnitude star, it is in principle visible just above the horizon; I will however allow in what follows an altitude of ½ ° (actually the horizon at the site estimated via satellite images looks flat towards the south-east). Then, it can be seen that the above azimuths match the rising azimuths of Sirius in the following approximate dates:[1]

Structure D    172°   9100 BC
Structure C    165°   8750 BC
Structure B    159°   8300 BC

3. Conclusions

The above arguments suggest that the structures of Gobekli Tepe were conceived to celebrate, and then follow in the course of the centuries, the appearance of a brilliant "guest" star in the sky: Sirius. Of course, although being fascinating, the hypothesis must be taken with due caution; in particular, the relative chronology between the structures is unclear (Dietrich 2013). Getting more insight in the symbolic world of the builders would certainly be of help; many of the animals are indeed tempting as representation of constellations (Belmonte and García 2010), and – curiously enough – one of the most elaborated stelae presents an upper register with three "bags" which are pretty similar to the three "houses of the sky" occurring in the much (very much!) later Babylonian "kudurru" traditions. Similar analyses in other PPN sites with megalithic structures would also be of help; further, recently discovered, inter-visible sites seem to align along a north-south direction (Guler, Celik and Guler 2012) .

---

1   The program used for the simulations (Starry Night Pro) takes into account the proper motion of Sirius.

As a final observation, it should be noted that a further structure uncovered at Gobekli Tepe (labelled F) has an estimated azimuth of 59° (if it was open to the north-east, as it seems) which is pretty close to that of the rising sun at the summer solstice. On the pillar 43 of enclosure D a suggestive, unique scene is represented: a sort of vulture with human traits delicately "rises up" with a wing what seems to be a sphere, or a disk. May this be a representation of the Heliacal rising of the newly born star we today call Sirius, which – as can be easily verified - occurred just a few days before the summer solstice at the end of the 10$^{th}$ millennium BC at the latitude of Gobekli Tepe?

**References**

Belmonte, J. and García, C. 2010, Astronomy, Landscape and Power in Eastern Anatolia
Proc. of SEAC 2010, Gilching.

Celik, B. 2001 Karahan Tepe: a new cultural centre in the Urfa area in Turkey
Documenta Praehistorica XXXVIII

Collins, A. 2013 Gobekli Tepe: its cosmic blueprint revealed.
http://www.andrewcollins.com/page/articles/Gobekli.htm

Dietrich, O. 2013 Gobekli Tepe. In PPND - The platform for neolithic radiocarbon dates
http://www.exoriente.org/associated_projects/ppnd_site.php?s=25

Güler M., Celik, B. Guler, G. 2012 New pre-pottery neolithic settlements from Viransehir district. Anadolu / Anatolia 38, 2012

Hoskin, M. 2001. Tombs, temples and their orientations, Ocarina books, Bognor Regis.

Magli, G. 2009. Mysteries and Discoveries of Archaeoastronomy, Springer-Verlag, NY

Ruggles, C. L. N. (2005) Ancient Astronomy: An Encyclopedia of Cosmologies and Myth ABC-CLIO, London.

Schmidt K. 1998. Beyond Daily Bread: Evidence of Early Neolithic Ritual from Gobekli Tepe. Neo-Lithics 2(98): 1–5.

Schmidt K..2001. Göbekli Tepe, Southeastern Turkey. A preliminary report on the 1995–1999 excavations. Paléorient 26(1): 45–54.

Schmidt K. 2006. Sie bauten die ersten Tempel. Das rätselhafte Heiligtum der Steinzeitjäger. Die archäologische Entdeckung am Göbekli Tepe. München: C.H. Beck

Schmidt K. 2010. Göbekli Tepe: the Stone Age Sanctuaries. New result of ongoing excavations with a special focus on sculptures and high reliefs. In M. Budja (ed.), 17th Neolithic Studies. Documenta Praehistorica 17: 239–256.

Schoch, R. 2012 Forgotten Civilization: The Role of Solar Outbursts in Our Past and Future by Inner Traditions, NY.

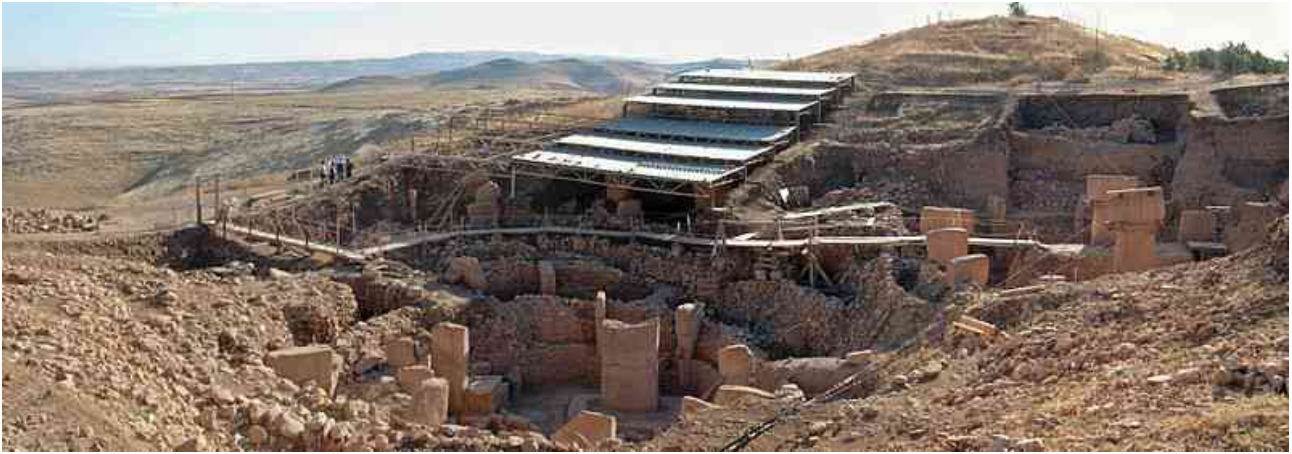

**Fig.1**
**Gobekli Tepe. The megalithic enclosures, view from the south (image in the public domain)**

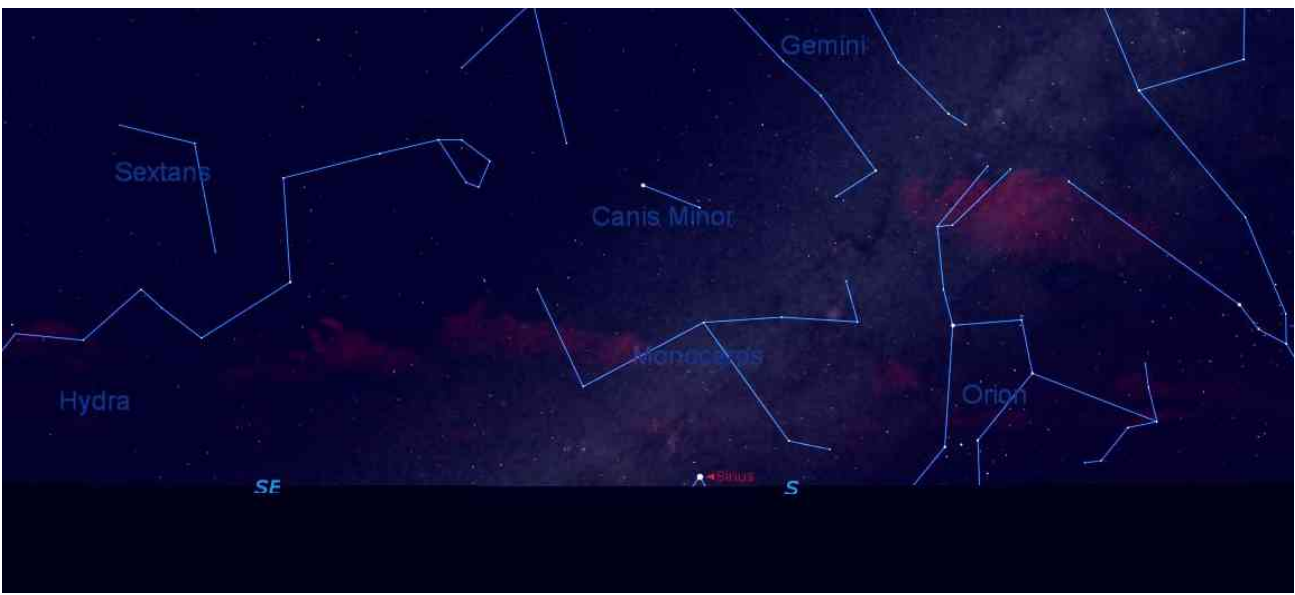

**Fig.2**
**Gobekli Tepe, 9100 BC. Rising of Sirius at azimuth 172°, a few days before summer solstice.**